\documentclass[sn-mathphys-num]{sn-jnl}
\usepackage{amsmath,amsfonts,amssymb}
\usepackage{algorithmic}
\usepackage{algorithm}
\usepackage{array}

\usepackage[caption=false,font=normalsize,labelfont=sf,textfont=sf]{subfig}
\usepackage{textcomp}
\usepackage{stfloats}
\usepackage{url}
\usepackage{verbatim}
\usepackage{graphicx}
\usepackage{booktabs}
\usepackage{multirow}
\usepackage{tabularx}
\usepackage{hyperref}
\usepackage{tikz}
\usetikzlibrary{positioning, arrows.meta, shapes.geometric}
\usepackage{comment}

\makeatletter
\renewcommand{\@maketitle}{%
    \null\vskip6pt%
    \hsize\textwidth\parindent0pt%
    {\hbox to \textwidth{{\Artcatfont\ArtType\hfill}\par}}
    \ifx\@title\empty\else%
        \removelastskip\vskip10pt\nointerlineskip%
        {\Titlefont\@title\par}
    \fi%
    \ifx\@subtitle\empty\else%
        \vskip4pt%
        {{\SubTitlefont\@subtitle\par}}
    \fi%
    \ifnum\aucount>0
        \global\punctcount\aucount%
        \vskip10pt%
        \artauthors\par%%     authors and emails
        {\vskip4pt\addressfont\auaddress\par%%      corresponding address
	 \removelastskip\vskip8pt%
	\ifnum\emailcnt>0\relax%
           \ifx\corrauthemail\@empty\else{\ifnum\aucount>1*\fi}%
	   Corresponding author(s). E-mail(s): \corrauthemail\par\fi%
	   \ifx\authemail\@empty\else Contributing authors:\ \authemail\fi%
        \fi%
        \ifequalcont{\par$^{\dagger}$\@equalconttext\par}\fi%
	 \removelastskip\vskip8pt%
        \ifpresentaddress{\par\@presentaddresstext\par}\fi%
	}
     \fi%
     {\printabstract\par}%
     {\printkeywords\par}%
     \ifx\@pacs\empty\else%
       \loop\ifnum\PacsCount>0%
          \csname\romannumeral\PacsTmpCnt StorePacsTxt\endcsname\par%
          \StepDownCounter{\PacsCount}%
          \StepUpCounter{\PacsTmpCnt}%
       \repeat%
    \fi%
    \removelastskip\vskip14pt\vskip0pt}%
\makeatother

\begin{document}

\title{ST-GAT: A Leakage-Free Spatio-Temporal Graph Attention Framework for Robust Network Intrusion Detection}

\author[1]{\fnm{Nikhil} \sur{Parkar}}
\author[1]{\fnm{Sanskar} \sur{Pandey}}
\author[2]{\fnm{Siddhi} \sur{Prajapati}}
\author*[1]{\fnm{Suganeshwari} \sur{G}}\email{suganeshwari.g@vit.ac.in}
\author[3]{\fnm{Syed} \sur{Ibrahim}}

\affil[1]{\orgdiv{School of Computer Science and Engineering}, \orgname{VIT Chennai}}
\affil[2]{\orgname{University of California, Irvine}}
\affil[3]{\orgname{VIT Mauritius}}

% The paper headers

\abstract{Intrusion detection systems are a central defence in modern networks. Deep learning models report strong anomaly-detection results, but evaluations on standard benchmarks such as CSE-CIC-IDS-2018 are often inflated by two methodological problems: temporal leakage from overlapping sliding windows, and contamination from preprocessing fitted before the train/test split. Per-flow classifiers also ignore the spatio-temporal dependencies across sequences of network flows. We propose a Spatio-Temporal Graph Attention Network (ST-GAT) that represents traffic as a sequence of graph snapshots. It builds $k$-Nearest Neighbour ($k$-NN) edges within each window, applies multi-head Graph Attention Network (GAT) layers for spatial aggregation, and uses a Bidirectional Long Short-Term Memory (BiLSTM) network with temporal attention for sequence modelling. To prevent leakage, we use non-overlapping temporal windows (stride equal to window length) and fit all preprocessing on the training split only. We compare three class-imbalance strategies: inverse class-frequency weighting, dynamic balanced sampling, and few-shot metric prototyping. Across three random seeds, ST-GAT with inverse class weighting reached a Macro-F1 of 96.65\% ($\pm$0.87) and an accuracy of 99.14\%. This is 14.3 percentage points higher in Macro-F1 than a per-flow Random Forest baseline (82.34\%). Balanced sampling reduced Macro-F1 by 2.3 points (to 94.37\%) relative to loss weighting. This suggests that aggressive resampling of small, noisy minority classes can hurt detection in heavily imbalanced traffic.}


\maketitle

\section{Introduction}\label{sec:intro}

As more devices connect over high-speed networks, the attack surface grows, and cyberattacks have become more frequent and more sophisticated. A network intrusion detection system (NIDS) monitors traffic to detect intrusion attempts as they occur. Per-flow classifiers such as Random Forest and Multi-Layer Perceptron are common NIDS baselines. However, they treat each flow independently, so they cannot represent attacks that are spread across many flows, such as botnet coordination and slow-rate denial of service. This limitation motivated recent graph-based detectors~\cite{Bondar2025, Rekik2025, Wang2025, Ghosh2025}.

Recent studies therefore use deep models. Graph Neural Networks (GNNs) model relationships among concurrent flows (the spatial aspect), and recurrent networks model how traffic evolves over time (the temporal aspect). On benchmarks such as CSE-CIC-IDS-2018, however, reported results are often hard to compare because of two recurring evaluation errors. The first is pre-split contamination: feature scaling or selection is fitted on the full dataset before the train/test split. The second is temporal leakage: overlapping sliding windows place the same flows in both training and test samples. At the same time, the heavy class imbalance of real traffic encourages the use of synthetic oversampling, often without checking whether it helps on the rarest, noisiest classes.

Most existing detectors model temporal dependencies or spatial relationships, but not both. Few combine joint spatio-temporal modelling with an evaluation protocol designed to rule out leakage. This gap motivates the ST-GAT framework.

ST-GAT treats network traffic as a sequence of discrete graph snapshots. Each individual network flow corresponds to a single graph node, and for each non-overlapping window of $T = 20$ flows, exactly one graph snapshot is generated with intra-window $k$-Nearest Neighbour ($k$-NN) similarity edges. Graph attention layers then learn how much weight each neighbour should receive. Finally, a BiLSTM with multi-head temporal attention models the order of flows across the window.

Our pipeline uses disjoint (non-overlapping) time windows and fits all preprocessing parameters on the training split only. We evaluate ST-GAT on CSE-CIC-IDS-2018 and compare three class-imbalance strategies. In our experiments, simple inverse class weighting outperformed both balanced resampling and few-shot prototyping. The contributions of this paper are as follows:

\begin{itemize}

    \item We propose ST-GAT, which combines multi-head graph attention over $k$-NN flow graphs with a BiLSTM and temporal attention. On CSE-CIC-IDS-2018 it reaches a mean Macro-F1 of 96.65\% and a mean accuracy of 99.14\% over three seeds.
    \item We describe a preprocessing and windowing pipeline designed to prevent the two leakage modes above. Scalers and variance thresholds are fitted on the training split only, and windows use a stride equal to their length, so that no flow appears in more than one window.
    \item We compare inverse class-frequency weighting with dynamic balanced sampling and few-shot metric prototyping on the same architecture. Inverse weighting achieved the highest Macro-F1, and balanced sampling the lowest, among the deep variants.
\end{itemize}

The rest of the paper is organised as follows. Section~\ref{sec:related} reviews classical machine learning and graph-based deep learning approaches to network intrusion detection. Section~\ref{sec:method} describes the methodology: data preprocessing, exploratory data analysis (EDA), spatio-temporal graph construction, the ST-GAT architecture, and the class-imbalance strategies we evaluate. Section~\ref{sec:setup} gives the experimental setup, including hyperparameters and evaluation metrics. Section~\ref{sec:results} presents the results, the comparison of balancing strategies, and the ablation study. Section~\ref{sec:conclusion} concludes and outlines future work.

\section{Related Work}\label{sec:related}
% ============================================================

We group prior work into four areas: (A) classical and ensemble machine learning baselines, (B) deep sequential models, (C) graph-based and spatio-temporal models, and (D) work on data leakage and evaluation methodology. Table~\ref{tab:related_summary} at the end of this section compares the most closely related works with the proposed framework.

% ------------------------------------------------------------
\subsection{Classical and Ensemble Machine Learning Baselines}
% ------------------------------------------------------------

Early NIDS research used shallow learners on the KDD Cup~1999 and NSL-KDD datasets. These benchmarks have well-known problems, including redundant records, unrealistic class distributions, and no encrypted traffic. To address them, the Canadian Institute for Cybersecurity released CIC-IDS2017~\cite{Sharafaldin2018} and then CSE-CIC-IDS-2018~\cite{CSECICIDS2018}.

CSE-CIC-IDS-2018 contains ten days of traffic. It was generated by an attacking infrastructure of 50 machines against a victim organisation of 420 machines and 30 servers~\cite{CSECICIDS2018}. The processed release contains over 16 million flows labelled with 14 attack types, including Botnet, Infiltration, several DoS and DDoS variants, brute-force, and web attacks. The dataset is more realistic than its predecessors, but it contains labelling errors and feature-extraction artefacts introduced by CICFlowMeter~\cite{Liu2022}.

\subsubsection{Tree Ensemble and Gradient Boosting Approaches}

Random Forest (RF) and gradient-boosted trees are the strongest classical baselines on CSE-CIC-IDS-2018. Hagar and Gawali~\cite{Hagar2022} trained three separate models (a CNN, an LSTM, and an Apache Spark classifier) on all 15 classes of the dataset. They used Random Forest to select 19 of 84 features and applied over- and undersampling to reduce the imbalance. They report F1-scores of 1.00 for most classes with the Spark model.
The paper does not report that normalisation and resampling were fitted on the training portion only. Preprocessing fitted before the split exposes the model to statistics of the test distribution that would not be available at deployment. This is \emph{pre-split data contamination}, and it can make reported metrics optimistic.

Ahmad et al.~\cite{Ahmad2021} systematically reviewed ML and DL approaches to NIDS. They reported that high accuracy is common in work based on CIC datasets, but few studies document how they prevent preprocessing errors.

Gradient-boosting methods such as XGBoost achieve high aggregate accuracy on CSE-CIC-IDS-2018. They handle missing values natively, regularise through shrinkage and column subsampling, and suit the tabular statistics that CICFlowMeter produces (for example \texttt{Flow Duration}, \texttt{Fwd Packet Length Mean} and \texttt{Bwd IAT Std}). However, these are per-flow, stateless classifiers. They treat each flow as an independent, exchangeable sample and cannot represent the \emph{temporal ordering} of flows. Slow-rate DoS and APT lateral movement unfold over many consecutive flows, so a model with no notion of sequence cannot capture them as multi-step patterns. This limitation led to the use of sequential deep models.

Liu et al.~\cite{Liu2022} audited the full creation process of CIC-IDS-2017 and CSE-CIC-IDS-2018. They found previously undocumented errors in attack orchestration, CICFlowMeter feature generation, documentation, and labelling. They also released a corrected, re-labelled version of the dataset.
Liu et al.\ argue that these errors undermine many published results, because a classifier can exploit artefacts of mislabelled or malformed flows instead of learning the attack behaviour itself. In our pipeline, we remove flows with infinite or NaN feature values before the split (Section~\ref{sec:preproc}).

% ------------------------------------------------------------
\subsection{Deep Sequential Models: RNNs, LSTMs, and Attention}
% ------------------------------------------------------------

Because stateless classifiers cannot use temporal context, NIDS research moved to recurrent architectures. Long Short-Term Memory (LSTM)~\cite{Hochreiter1997} uses gating to mitigate the vanishing gradients of simple RNNs, and it was among the first deep temporal models applied to NIDS.

\subsubsection{Unidirectional and Bidirectional LSTM Variants}

Gao~\cite{Gao2022} combined a 1D convolutional network, a Bidirectional LSTM (BiLSTM), and an attention mechanism for intrusion detection in cloud environments. The convolutional layers extract local patterns from each flow's $d$-dimensional feature vector using $K$ filters of width $w$:

\begin{equation}
\mathbf{h}_t = \text{ReLU}(\mathbf{W}_c \star \mathbf{x}_{t} + \mathbf{b}_c)
\end{equation}
These features feed a BiLSTM, which concatenates its forward and backward hidden states:
\begin{equation}
\overrightarrow{\mathbf{h}}_t \oplus \overleftarrow{\mathbf{h}}_t
\end{equation}

On a cleaned and mapped version of KDD Cup~1999, the model reached an average accuracy of 95.50\% with a false-positive rate of 4.24\%~\cite{Gao2022}. KDD Cup~1999 predates modern attack traffic and has the redundancy problems noted above.

Dai et al.~\cite{Dai2024} proposed a CNN--BiLSTM--attention model. A CNN first projects the input features to a $d'$-dimensional representation. BiLSTM layers then accumulate temporal context, and an attention layer weights the resulting states:

\begin{equation}
\text{Attention}(\mathbf{Q},\mathbf{K},\mathbf{V}) = \text{softmax}\!\left(\frac{\mathbf{Q}\mathbf{K}^\top}{\sqrt{d_k}}\right)\mathbf{V}
\end{equation}

The model was evaluated on NSL-KDD, UNSW-NB15, and CIC-DDoS2019. To handle class imbalance, it uses the Equalization Loss~v2 objective rather than resampling.
Where scaling statistics are computed on the full dataset, the means and variances are pulled slightly toward the test distribution. The effect is largest for classes with few samples.

\subsubsection{Transformer-Based Sequential Models}

Transformer architectures~\cite{Vaswani2017} have also been adapted to NIDS. Zhang et al.~\cite{Li2025} combined a Transformer branch with CNN--BiLSTM branches for IoT intrusion detection, removed noise with Isolation Forest and LOF, and handled class imbalance with Borderline-SMOTE. They report 99.80\% accuracy on CIC-IDS2017 and 97.95\% on BoT-IoT. The Transformer's self-attention lets every position attend to every other, at $O(n^2)$ memory cost in sequence length. The paper does not state the sliding-window stride. Nor does it state whether Borderline-SMOTE was applied before or after the train/test split. Both details determine whether the training and test sets share raw flows or synthetic samples derived from test flows. Such omissions are common in the BiLSTM and Transformer NIDS literature and make independent reproduction difficult.
% (Citation key Li2025 retained; the verified first author is Chunhui Zhang.
% The bibliography entry has been corrected.)

Wang et al.~\cite{ResTranBiLSTM2023} proposed Res-TranBiLSTM, which combines residual connections, Transformer self-attention and BiLSTMs for IoT intrusion detection. They evaluated it on NSL-KDD, CIC-IDS2017, and MQTTset, and used SMOTE-ENN to reduce class imbalance. It was not evaluated on CSE-CIC-IDS-2018. Our experiments test whether synthetic oversampling helps on the noisy minority classes of that dataset.

\subsubsection{Summary of Sequential Model Limitations}

Three methodological problems recur in the BiLSTM and attention literature. \emph{First}, overlapping sliding windows: with window length $T$ and stride $s<T$, consecutive windows share $T-s$ raw flows. Without a gap between the end of the training data and the start of the test data, test flows also appear in training windows.
\emph{Second}, scaling before partitioning: fitting a feature scaler on the full corpus before the split passes information about the test distribution into the feature space. \emph{Third}, synthetic oversampling (SMOTE and its variants) is often used without checking that the synthetic samples represent real minority-class behaviour in noisy datasets.

ST-GAT addresses these as follows. It uses non-overlapping windows of $T = 20$ flows with stride $s = T$. It fits scalers and feature-selection steps on the training split only. It handles class imbalance by weighting the cross-entropy loss with inverse class frequencies instead of generating synthetic samples.

% ------------------------------------------------------------
\subsection{Graph Neural Networks for Intrusion Detection}
% ------------------------------------------------------------

Classical and sequential NIDS models assume that flows are either \emph{independent} or \emph{linearly ordered}. They ignore the \emph{relations} between flows. Flows can share a source and destination, originate from the same IP address, or have similar statistics while occurring at different times. None of these relations is captured by a single feature vector or a 1-D sequence. Graph Neural Networks (GNNs)~\cite{Wu2021} operate on graph-structured data and propagate features along edges that encode such relationships.

\subsubsection{Graph Convolutional Networks}

The Graph Convolutional Network (GCN) of Kipf and Welling~\cite{Kipf2017}
defines the layer-wise propagation rule

\begin{equation}
\mathbf{H}^{(l+1)} = \sigma\!\left(\tilde{\mathbf{D}}^{-1/2} \tilde{\mathbf{A}}\tilde{\mathbf{D}}^{-1/2}\mathbf{H}^{(l)}\mathbf{W}^{(l)}\right)
\end{equation}

where $\tilde{\mathbf{A}} = \mathbf{A} + \mathbf{I}$ is the adjacency matrix with self-loops, $\tilde{\mathbf{D}}$ is its degree matrix, and $\mathbf{W}^{(l)}$ is the trainable weight matrix of layer $l$. The symmetric normalisation keeps the propagation numerically stable. However, aggregation weights are fixed by node degrees and do not depend on node features, so every neighbour of a given degree contributes equally regardless of relevance. In NIDS, this means a benign background flow can influence neighbourhood aggregation as much as an attack flow, which dilutes the discriminative signal.

\subsubsection{Graph Attention Networks}

The Graph Attention Network (GAT) of Veli\v{c}kovi\'{c} et al.~\cite{Velickovic2018} replaces fixed weights with a learned attention coefficient for each edge $(i, j)$:

\begin{equation}
\alpha_{ij} = \frac{\exp\!\left(\text{LeakyReLU}\!\left(\mathbf{a}^\top [\mathbf{W}\mathbf{h}_i \| \mathbf{W}\mathbf{h}_j]\right)\right)}{\sum_{k \in \mathcal{N}(i)}\exp\!\left(\text{LeakyReLU}\!\left(\mathbf{a}^\top [\mathbf{W}\mathbf{h}_i \| \mathbf{W}\mathbf{h}_k]\right)\right)}
\end{equation}

where $\|$ denotes concatenation and $\mathbf{a}$ is a learnable attention vector. Multi-head GAT computes $K$ independent attention heads and concatenates (or averages) their outputs, which stabilises training~\cite{Velickovic2018}:

\begin{equation}
\mathbf{h}'_i = \Big\|_{k=1}^{K} \sigma\!\left(\sum_{j \in \mathcal{N}(i)} \alpha_{ij}^k \mathbf{W}^k \mathbf{h}_j\right)
\end{equation}

ST-GAT uses multi-head GAT as its spatial feature extractor.

Lo et al.~\cite{Lo2022} proposed E-GraphSAGE, an early GNN-based flow NIDS. It treats hosts as nodes and flow records as edges, so it can use both edge features and graph topology. It outperformed prior methods on four NIDS benchmarks. However, building such host graphs requires source and destination endpoint identifiers, and a model that depends on them cannot run where those fields are withheld or anonymised. ST-GAT does not use endpoint identifiers. It builds graphs with $k$-Nearest Neighbours ($k$-NN) over the standardised flow-feature space, using only the per-flow statistics that CICFlowMeter-processed datasets provide. Ngo et al.~\cite{Ngo2025} similarly built Top-$K$ attribute-similarity graphs instead of physical topology. Our construction differs in being computed separately within each temporal window. Each node $v_i$ is connected to its $k$ nearest neighbours by Euclidean distance in feature space:

\begin{equation}
\mathcal{E}_{kNN} = \{(v_i, v_j) \mid v_j \in \text{kNN}(v_i)\}
\end{equation}

Similarity-based edges can link behaviourally similar flows, for example two Botnet flows with similar packet-size distributions, that a host-topology graph would not connect.

\subsubsection{Behaviour-Similarity and Hybrid GNN Models}

Wang et al.~\cite{Wang2025} proposed BS-GAT, a graph attention network whose edges come from behaviour rules shared between flows rather than from network topology. This improves graph construction, but the behaviour rules must be designed by hand. Jahin et al.~\cite{CAGNGAT2025} proposed CAGN-GAT Fusion, which combines contrastive learning with adaptive, threshold-based graph construction and uses edge perturbation and feature masking for augmentation. They evaluated it on KDD Cup~1999, NSL-KDD, UNSW-NB15, and CIC-IDS2017 and report an F1 of 0.946 on CIC-IDS2017 without augmentation. They build the graph after an 80:20 train/test split. Like BS-GAT, it builds graphs from individual flows but does not model how traffic evolves across time windows.

Other recent GNN detectors have explored edge-aware and explainable GraphSAGE variants~\cite{Lusa2025}, multi-hop attention over edge features~\cite{Deng2025}, residual attention GNNs~\cite{Zhang2025}, self-supervised learning on flow graphs~\cite{Xu2024}, and federated GNNs~\cite{Ghosh2025}. Comparative studies report that GAT often outperforms GCN and GraphSAGE on IoT traffic~\cite{Bondar2025, Saxena2025}. A smaller group of methods combines graphs with temporal dynamics: neural ODEs over changing topologies~\cite{Rekik2025}, dynamic-graph models~\cite{Liu2025}, and sliding-window CNN--GRU models enhanced with GNNs~\cite{Chinnasamy2025}. Most GNN detectors, however, treat each graph as an independent sample with no temporal link to the next window, whereas BiLSTM models ignore relationships between flows within a window. ST-GAT combines a multi-head graph attention layer (spatial module) with a BiLSTM and multi-head temporal attention (temporal module) in a single model.

\subsubsection{Spatio-Temporal Architectures in Adjacent Domains}

Spatio-temporal GNNs are well developed in road-traffic forecasting and in anomaly detection for industrial control systems (ICSs). Presekal et al.~\cite{Musleh2024} combined enhanced graph convolution with an LSTM (GC-LSTM) to detect and correlate advanced persistent threats in cyber-physical power systems. Koistinen et al.~\cite{STAGNN2025} used dynamically learned graphs with attention for unsupervised anomaly detection on the SWaT water-treatment testbed. These architectures inspired the design of ST-GAT. However, they operate on known sensor and device graphs with time-series data. Applying the idea to a tabular, multi-class, highly imbalanced dataset such as CSE-CIC-IDS-2018 requires three changes: a topology-inference step, a supervised classification head, and explicit imbalance handling.
% (Keys Musleh2024 and STAGNN2025 retained; the bibliography author lists
% were wrong and have been corrected to the verified authors. See the
% bibliography.)

% ------------------------------------------------------------
\subsection{Methodological Critique: Data Leakage, Evaluation Protocols,
            and Class-Imbalance Treatment}
% ------------------------------------------------------------

A growing body of work re-examines how NIDS and other ML benchmarks are evaluated. It argues that flawed protocols lead to systematic overestimation of real-world performance~\cite{kapoor2023leakage, Liu2022, Bouke2023}.

\subsubsection{Temporal Leakage via Overlapping Windows}

Temporal leakage occurs when a model is trained on information that comes from after the observations it is tested on. In NIDS, it usually arises from sliding windows with stride $s < T$. A window $w_i = [f_t, \ldots, f_{t+T-1}]$ and the next window $w_j = [f_{t+s}, \ldots, f_{t+s+T-1}]$ share $T - s$ flows. If the train/test boundary falls inside that overlap, the same flows appear in both sets. Kapoor and Narayanan~\cite{kapoor2023leakage} survey leakage across 17 scientific fields. Their taxonomy includes temporal leakage and non-independence between training and test samples. They show that such leakage can reverse conclusions about which model performs best. For rare classes, even a few flows duplicated across the boundary can noticeably change the measured recall. ST-GAT avoids this form of leakage by construction: windows of length $T = 20$ use stride $s = T = 20$, so no flow appears in more than one window.

\subsubsection{Pre-Split Data Contamination}

Any data-dependent preprocessing done before the train/test split is pre-split contamination. Bouke and Abdullah~\cite{Bouke2023} measured this directly. They compared six classifiers (Decision Tree, Gradient Boosting, $k$-NN, SVM, Random Forest and Logistic Regression) on NSL-KDD, UNSW-NB15, and KDD Cup~1999, using preprocessing pipelines with and without leakage. The leaky pipelines gave higher, but unreliable, accuracy, and sensitivity to leakage varied by algorithm. For rare attack classes, a small change in aggregate metrics can correspond to many missed intrusions per evaluation period. In ST-GAT, the feature scaler and feature selectors are fitted on the training split only. Because $k$-NN edges are computed inside each 20-flow window (Section~\ref{sec:graph}), graph construction uses no information from outside the window.

\subsubsection{Class Imbalance: Oversampling vs. Loss-Weighting}

NIDS datasets are highly imbalanced. In the subset of CSE-CIC-IDS-2018 used here, \texttt{Benign} makes up 1,367 of the 1,855 test windows. The rarest retained classes, \texttt{DoS attacks-Slowloris} and \texttt{DoS attacks-GoldenEye}, have 3 and 9 test windows respectively (Table~\ref{tab:per_class_f1}). The most widely used remedy is synthetic oversampling with SMOTE~\cite{Chawla2002} and its variants. More recent work generates minority samples with GANs and VAE--GAN hybrids~\cite{Yang2023, Li2025new, Kang2025} or uses minority-focused ensembles~\cite{Oh2025}. Comparative evaluations report mixed benefits across datasets~\cite{Shanmugam2024, Imbalance2026}.

SMOTE creates synthetic samples by linear interpolation between a minority sample and one of its nearest minority-class neighbours~\cite{Chawla2002}. When a minority class forms a small, noisy cluster in feature space, interpolation can produce near-duplicates, which encourage overfitting, or points that fall inside other classes' regions, which act as label noise. The classifier may then overfit the noisy minority region while misclassifying real minority samples near the class boundaries. We refer to this effect as \textit{noisy minority over-penalisation}.

The alternative, inverse class-frequency weighting, gives each class $c$ a weight in the cross-entropy loss:
\begin{equation}
w_c = \frac{N}{C \cdot N_c}
\end{equation}

where $N$ is the total number of samples, $C$ is the number of classes, and $N_c$ is the number of samples in class $c$. This method neither changes the feature space nor creates synthetic samples. Instead, it scales each sample's gradient contribution according to how rare its class is, so errors on rare classes cost more during training. In our experiments (Section~\ref{sec:imbalance_results}), this weighting gave a mean Macro-F1 of 96.65\% on the held-out test split. This was higher than both balanced resampling (94.37\%) and few-shot prototyping (95.43\%) on the same GAT backbone.

Dinh et al.~\cite{Imbalance2026} reach a consistent conclusion on other benchmarks. They compared traditional and deep-learning oversampling methods on UNSW-NB15 and TON\_IoT. On UNSW-NB15, severe class imbalance and class overlap limited the gains from resampling, while on TON\_IoT classifiers performed well even without oversampling.

\subsubsection{Comprehensive Surveys and Open Problems}

Ahmad et al.~\cite{Ahmad2021} surveyed ML and DL approaches to network intrusion detection and identified recurring open challenges. These include the lack of temporal modelling in flow-level classifiers, the lack of relational modelling between flows, inconsistent evaluation protocols that prevent comparison across studies, and the difficulty of detecting very rare attack categories. ST-GAT targets the first two directly, while our evaluation protocol and empirical comparison of imbalance strategies analyze the last two on a single benchmark.

Wu et al.~\cite{Wu2021} survey the main families of GNN architectures (recurrent, convolutional, graph autoencoders and spatio-temporal GNNs) and their applications. We chose multi-head GAT over GCN because learned, feature-dependent attention can down-weight irrelevant neighbours. This is useful when $k$-NN neighbourhoods mix benign and attack flows. Our ablation (Section~\ref{sec:ablation}) and the GAT--GCN comparison (Table~\ref{tab:main_results}) test this choice.

Table~\ref{tab:related_summary} shows that the most closely related works fall into three groups: (a) graph-based models that learn spatial features but do not model time, (b) sequential BiLSTM or attention models that ignore relations between flows and whose evaluations are exposed to data leakage, and (c) evaluation and dataset audits that identify these problems without proposing an architecture that addresses them together.

In summary, NIDS architectures have moved from rule-based expert systems through classical ML and ensembles to deep sequential models and, most recently, graph-based and spatio-temporal networks. In parallel, a growing literature on \emph{methodological rigour} has shown how temporal leakage, pre-split contamination, and poor handling of class imbalance inflate benchmark results. ST-GAT draws on both lines of work. It combines graph-based spatial modelling with sequential modelling and is evaluated under a protocol designed to avoid the first two of these flaws.

% ============================================================
%  Summary Table (two-column spanning)
% ============================================================

\begin{table*}[htpb]
\renewcommand{\arraystretch}{1.4}
\caption{Summary of Most Closely Related Works and Alignment with the Proposed ST-GAT Framework}
\label{tab:related_summary}
\centering
\small
\resizebox{\textwidth}{!}{
\begin{tabular}{%
  >{\raggedright\arraybackslash}p{2.6cm}  % Authors & Year
  >{\raggedright\arraybackslash}p{3.2cm}  % Key Contribution
  >{\raggedright\arraybackslash}p{2.6cm}  % Results
  >{\raggedright\arraybackslash}p{3.8cm}  % Opportunities (Flaws)
  >{\raggedright\arraybackslash}p{3.6cm}  % Alignment with ST-GAT
}
\toprule
\textbf{Authors \& Year} &
\textbf{Key Contribution} &
\textbf{Results / Metric Reported} &
\textbf{Opportunities for Extension (Methodological Flaws)} &
\textbf{Alignment with Proposed ST-GAT} \\
\midrule

Dai et al.~\cite{Dai2024} (2024) &
CNN + BiLSTM + attention for NIDS, with Equalization Loss v2. &
99.4\% Acc; 98.7\% Macro-F1 (NSL-KDD / UNSW-NB15 / CIC-DDoS2019). &
Scaling fitted before the split not ruled out; no graph-based spatial modelling. &
Adds $k$-NN spatial graphs, uses non-overlapping windows ($s=20$), and fits scaling on the training split only. \\

\midrule

Lo et al.~\cite{Lo2022} (2022) &
E-GraphSAGE: host-graph GNN with edge (flow) features. &
97.6\% Macro-F1. &
Requires endpoint identifiers to build host graphs; no temporal modelling across windows. &
Builds $k$-NN graphs from flow statistics alone and models temporal dependencies with BiLSTM and attention. \\

\midrule

Liu et al.~\cite{Liu2022} (2022) &
Audit of CIC-IDS-2017 and CSE-CIC-IDS-2018 errors; corrected re-release. &
Corrected labels reduce reported accuracy by up to 3\%. &
No new architecture proposed; does not address windowing or pre-split leakage. &
Removes malformed (infinite/NaN) flows and additionally prevents windowing and pre-split leakage. \\

\midrule

Hagar \& Gawali~\cite{Hagar2022} (2022) &
CNN, LSTM, and Apache Spark models with RF feature selection and resampling. &
Up to 100\% Acc; F1 = 1.00 for most classes. &
Resampling and normalisation not reported as train-only; no per-class Macro-F1 summary. &
Reports Macro-F1 (96.65\%) and per-class results under train-only preprocessing. \\

\midrule

Bouke \& Abdullah~\cite{Bouke2023} (2023) &
Measures the effect of preprocessing leakage on six ML classifiers. &
Leakage inflates accuracy by 1.5--4.0\%. &
Classical ML on tabular data only; windowing leakage not studied. &
Applies train-only preprocessing to a deep spatio-temporal model and removes window overlap. \\

\midrule

Appiahene et al.~\cite{Appiahene2026} (2026) &
GConvTrans (graph convolution + Transformer) on CSE-CIC-IDS-2018 (14 Feb.\ subset). &
96.94\% test Acc. &
Temporal-leakage controls not described; no comparison of imbalance strategies; F1 not reported. &
Uses non-overlapping windows and compares inverse class weighting (96.65\% Macro-F1) against sampling-based strategies. \\

\bottomrule
\end{tabular}
}
\end{table*}

\section{Methodology}\label{sec:method}
% ============================================================

ST-GAT is designed to capture spatio-temporal interactions between network flows without introducing data leakage. This section describes data preprocessing, exploratory analysis, spatio-temporal graph construction, the ST-GAT architecture, and the class-imbalance strategies we compare.

% ------------------------------------------------------------
\subsection{Dataset and Preprocessing}\label{sec:preproc}
% ------------------------------------------------------------

We evaluate our framework on CSE-CIC-IDS-2018, an enterprise-scale benchmark released collaboratively by the Communications Security Establishment (CSE) and the Canadian Institute for Cybersecurity (CIC)~\cite{CSECICIDS2018}. The dataset was generated over ten operational capture days (spanning February 14 to March 2, 2018) within a realistic cloud environment deployed on Amazon Web Services (AWS). 

The testbed architecture separates the network into attacking and victim infrastructures. The victim enterprise network consists of 420 client machines running various versions of Windows (7, 8.1, 10) and Linux (Ubuntu 14.04 and 16.04), alongside 30 servers distributed across five distinct departmental subnets. These servers host critical enterprise services, including Apache web servers, email exchange servers, domain controllers, Active Directory, and MySQL database management systems. The attacking infrastructure comprises 50 external AWS EC2 virtual machines distributed across four dedicated subnets executing coordinated, automated penetration routines.

Network traffic generation follows a dual-profile methodology:
\begin{itemize}
    \item \textbf{Benign Background Profiles (B-Profiles):} Normal traffic was generated by agents executing statistical models of human user behaviour. These B-profiles encapsulate natural usage patterns across HTTP, HTTPS, FTP, SSH, and email protocols, introducing realistic request-response sizes, transmission delays, and stochastic inter-arrival distributions.
    \item \textbf{Malicious Attack Profiles (M-Profiles):} Attack traffic covers seven major attack scenarios orchestrated according to a deterministic daily schedule:
    \begin{enumerate}
        \item \textit{Brute Force:} Automated credential stuffing and dictionary attacks against FTP and SSH services executed using Hydra.
        \item \textit{Heartbleed:} Memory-disclosure exploitation targeting the OpenSSL TLS heartbeat extension (CVE-2014-0160).
        \item \textit{Botnet:} Command-and-control (C\&C) interactions and malware propagation based on the Ares Python botnet framework.
        \item \textit{Denial of Service (DoS):} Application- and transport-layer resource exhaustion attacks, including DoS-Slowloris, DoS-SlowHTTPTest, DoS-Hulk, and DoS-GoldenEye.
        \item \textit{Distributed Denial of Service (DDoS):} High-volume distributed packet flooding attacks driven by LOIC (HTTP and UDP) and High Orbit Ion Cannon (HOIC).
        \item \textit{Web Attacks:} Injection and parameter tampering targeting web servers, including SQL Injection, Cross-Site Scripting (XSS), and Web-based Brute Force.
        \item \textit{Infiltration:} Multi-stage Advanced Persistent Threat (APT) simulation wherein a workstation inside the enterprise is compromised via a malicious PDF/Dropbox link, followed by internal network reconnaissance (Nmap) and lateral movement/privilege escalation via Metasploit.
    \end{enumerate}
\end{itemize}

Raw network packets were captured via network taps and converted into bidirectional flows using CICFlowMeter-V3~\cite{Sharafaldin2018, CSECICIDS2018}. Each flow is defined by the standard 5-tuple: $(\text{Source IP}, \text{Destination IP}, \text{Source Port}, \text{Destination Port}, \text{Protocol})$ bounded by a temporal flow-inactivity timeout. The resulting dataset produces 80 statistical flow features characterizing both forward (client-to-server) and backward (server-to-client) directions. These features encompass:
\begin{enumerate}
    \item \textit{Temporal and Cadence Statistics:} \texttt{Flow Duration}, flow inter-arrival times (\texttt{Flow IAT Mean}, \texttt{Std}, \texttt{Max}, \texttt{Min}), and directional inter-arrival times (\texttt{Fwd IAT Tot}, \texttt{Bwd IAT Tot}).
    \item \textit{Volume and Length Metrics:} Total packet counts (\texttt{Tot Fwd Pkts}, \texttt{Tot Bwd Pkts}), total payload lengths, and statistical distributions of packet lengths (\texttt{Packet Length Mean}, \texttt{Std}, \texttt{Variance}).
    \item \textit{Transmission Control and TCP Flags:} Frequency of TCP control flags (\texttt{FIN}, \texttt{SYN}, \texttt{RST}, \texttt{PSH}, \texttt{ACK}, \texttt{URG}, \texttt{CWR}, \texttt{ECE}), initial TCP window advertisement sizes (\texttt{Init Fwd Win Byts}, \texttt{Init Bwd Win Byts}), and segment size counts.
    \item \textit{Sub-flow and Activity Descriptors:} Sub-flow packet allocations and active/idle state duration statistics.
\end{enumerate}

Across the full 10-day capture, CSE-CIC-IDS-2018 comprises over 16 million flow records. In our evaluated corpus, the dataset exhibits a representative multi-class distribution with extreme long-tail class imbalance, ranging from majority benign traffic ($>$182,000 flows) to well-represented volumetric attacks (HOIC ~20,500; Hulk ~13,800), down to low-rate stealthy attacks such as DoS-Slowloris and Infiltration. The raw data also exhibit documented outliers, division-by-zero calculation anomalies, and embedded extraction artifacts~\cite{Liu2022}.

Preprocessing began with structural anomalies and undefined values. The dataset is a concatenation of several capture files, so repeated header rows (where \texttt{Label} equals the literal string ``Label'') were removed. All flow features were then cast to numeric types. Some features contained positive or negative infinity (\texttt{np.inf}, \texttt{-np.inf}) caused by division by zero in rate calculations. These values were replaced with \texttt{NaN}, and every flow containing a \texttt{NaN} was discarded.

Four classes have too few samples to populate the training, validation, and test splits: \texttt{SQL Injection}, \texttt{Brute Force -Web}, \texttt{Brute Force -XSS}, and \texttt{DDOS attack-LOIC-UDP}. These classes were removed. The evaluated capture corpus contains no samples of \texttt{DDoS attacks-LOIC-HTTP}, leaving 10 evaluated classes: \texttt{Benign} and 9 distinct attack categories.

Next, we removed uninformative and potentially misleading features. The metadata fields \texttt{Timestamp} and \texttt{Dst Port} were dropped so that the network would learn attack behaviour rather than memorise timing and port information. Four flag features (\texttt{Fwd PSH Flags}, \texttt{Bwd PSH Flags}, \texttt{Fwd URG Flags} and \texttt{Bwd URG Flags}) were dropped because they showed little or no variance. The rate features \texttt{Flow Byts/s} and \texttt{Flow Pkts/s} were also removed. They were the source of the infinite values described above and caused exploding gradients in preliminary training runs.

\texttt{Protocol} identifies the transport-layer protocol (TCP, UDP or ICMP) and is nominal. We cast it to a categorical string and one-hot encoded it, so the network does not treat protocol codes as ordinal values. The target \texttt{Label} was integer-encoded with a label encoder.

To avoid pre-split contamination, we partitioned the sequence windows before fitting any feature scalers. We employed a stratified split assigning 70\% of sequence windows to training, 15\% to validation, and 15\% to testing, preserving the empirical class proportions across all three partitions. Crucially, stratified splitting does not overcome or mitigate class imbalance itself; rather, it acts as an essential experimental safeguard that prevents partition mishaps (such as extreme minority attack classes with only a handful of examples being omitted entirely from either the training or test split), thereby ensuring consistent and representative evaluation support across all categories. Early stopping monitored validation Macro-F1 with a patience of 20 epochs, ensuring that the test split remained strictly held out until final evaluation.

Following feature selection via a Random Forest importance ranking fitted strictly on the training partition ($D = 35$ top features), a \texttt{MinMaxScaler} was fitted on the training split only and used to normalize features into $[0, 1]$ across training, validation, and test sets. As a result, no test-set statistics enter either feature selection or feature scaling, and distance computations during $k$-NN graph construction remain uniformly bounded across all dimensions.

% ------------------------------------------------------------
\subsection{Exploratory Data Analysis (EDA)}
% ------------------------------------------------------------

Before modelling, we analysed the preprocessed data to characterise the feature space and the degree of class imbalance. The findings informed the design of ST-GAT.

\subsubsection{Class Distribution and Imbalance}

\begin{figure}[htbp]
\centering
\includegraphics[width=\linewidth]{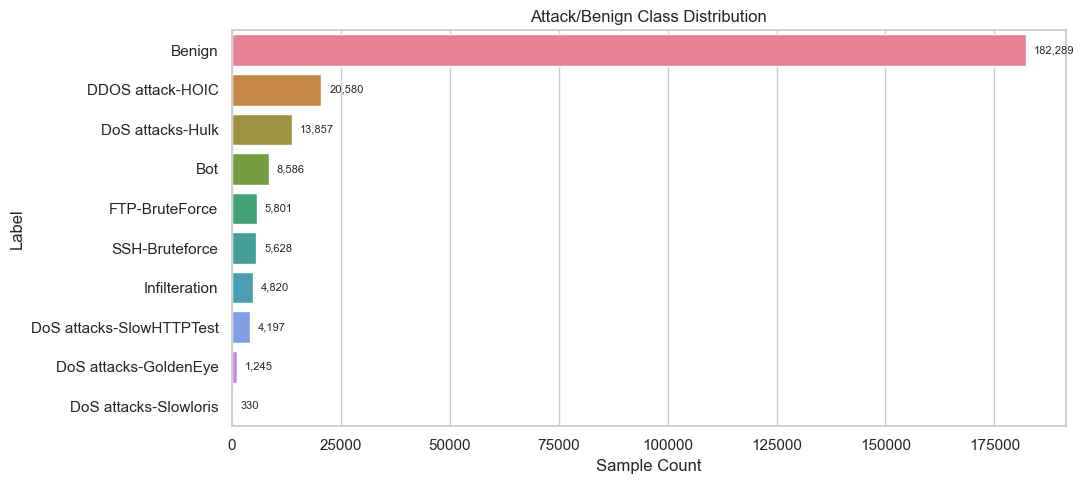}
\caption{Class distribution of the preprocessed data, showing a long-tailed imbalance between Benign and attack classes.}
\label{fig:class_dist}
\end{figure}

The class distribution is highly imbalanced, as is typical of real networks, where malicious traffic is a small fraction of total volume. As Fig.~\ref{fig:class_dist} shows, \texttt{Benign} dominates with more than 182,000 flows. Among attacks, \texttt{DDOS attack-HOIC} (about 20,500 flows) and \texttt{DoS attacks-Hulk} (about 13,800 flows) are well represented, while the other attack classes are much smaller.

Because of this imbalance, unweighted cross-entropy is a poor objective. A model trained with it can reach high accuracy by predicting \texttt{Benign} for every sample. In our test split, this trivial rule would already be correct for 73.7\% of windows (1,367 of 1,855) while detecting no attacks. This motivates the comparison of imbalance-handling strategies in Section~\ref{sec:imbalance}.

\subsubsection{Feature Variance and Multicollinearity}

\begin{figure}[htbp]
\centering
\includegraphics[width=\linewidth]{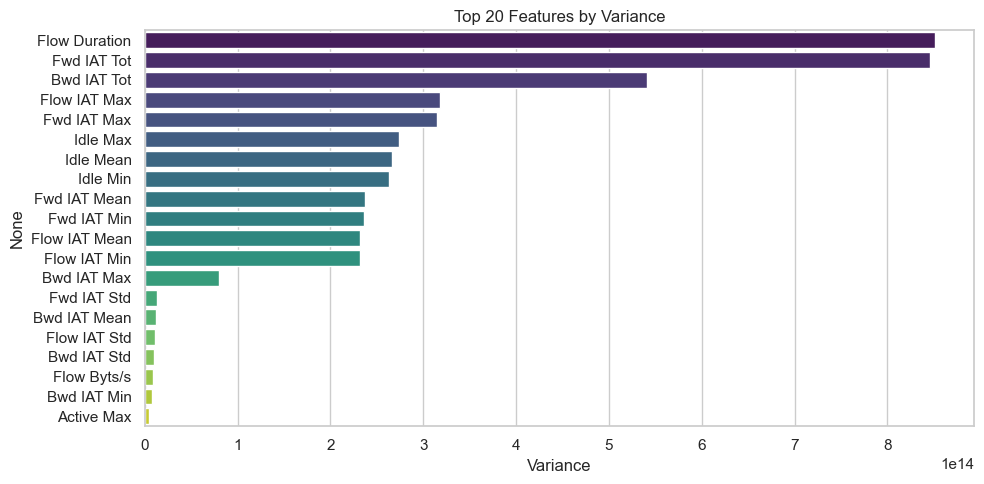}
\caption{Top 20 features ranked by variance, dominated by inter-arrival time (IAT) and flow-duration statistics.}
\label{fig:feature_var}
\end{figure}

Most of the variance in the (unscaled) data comes from a few timing features. Figure~\ref{fig:feature_var} shows that \texttt{Flow Duration}, \texttt{Fwd IAT Tot} and \texttt{Bwd IAT Tot} have variances up to about $10^{14}$. The prominence of inter-arrival time (IAT) features suggests that temporal structure carries much of the signal. We test this hypothesis in the ablation study (Section~\ref{sec:ablation}).

\begin{figure}[htbp]
\centering
\includegraphics[width=0.75\linewidth]{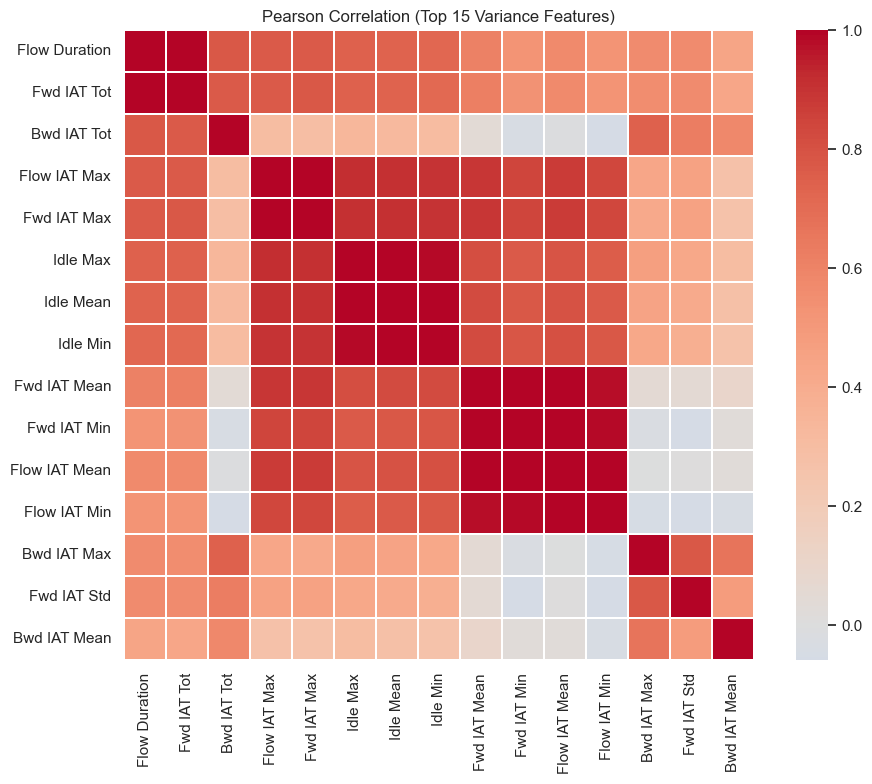}
\caption{Pearson correlation matrix of the 15 highest-variance features, showing clusters of strongly correlated features.}
\label{fig:correlation}
\end{figure}

The Pearson correlation matrix of the 15 highest-variance features (Fig.~\ref{fig:correlation}) shows strong multicollinearity. For example, \texttt{Flow IAT Max} and \texttt{Fwd IAT Max} are almost perfectly correlated ($r \approx 1.0$). Models that assume conditional independence of features, such as Naive Bayes, are poorly suited to such data. Neural models with dense and attention layers can instead learn latent representations that combine correlated inputs.

\subsubsection{Principal Component Analysis (PCA)}

\begin{figure}[htbp]
\centering
\includegraphics[width=0.75\linewidth]{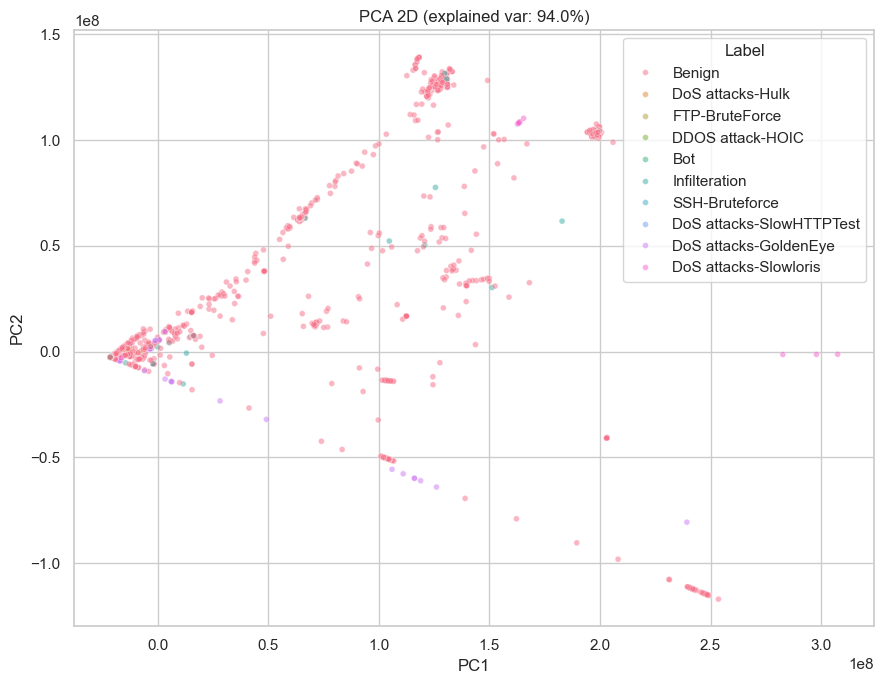}
\caption{2D principal component projection. Although the two components explain 94\% of the variance, classes overlap heavily, indicating that they are not linearly separable in this subspace.}
\label{fig:pca}
\end{figure}

We ran PCA to examine how linearly separable the classes are. As Fig.~\ref{fig:pca} shows, the first two principal components explain 94.0\% of the variance. Even so, the 2D projection shows heavy overlap between \texttt{Benign} and several attack classes.

In this linear projection, \texttt{Benign} samples surround the \texttt{DoS}, \texttt{Botnet} and \texttt{BruteForce} samples. The overlap indicates that these classes are not linearly separable in the leading principal components. This supports the use of a non-linear model such as ST-GAT, though it does not rule out other non-linear classifiers; the Random Forest baseline is also non-linear.

% ------------------------------------------------------------
\subsection{Spatio-Temporal Graph Construction}\label{sec:graph}
% ------------------------------------------------------------

NIDS datasets such as CSE-CIC-IDS-2018 are structured tabular collections where each row describes a single network flow $\mathbf{f}_i \in \mathbb{R}^D$. To process flow traffic with ST-GAT, we transform this continuous stream of flows into a discrete sequence of graph snapshots. In our formulation, each individual network flow corresponds to exactly one node in the graph. For every non-overlapping window of $T = 20$ flows, exactly one graph snapshot $\mathcal{G}_t = (\mathbf{X}_t, \mathbf{A}_t)$ is generated with an intra-window $20 \times 20$ adjacency matrix. This deterministic transformation proceeds in two stages: temporal window creation and spatial edge construction.

\subsubsection{Leakage-Free Temporal Window Generation}
The first stage groups contiguous flows into windows to capture temporal dynamics. Let $\mathcal{F} = \{\mathbf{f}_1, \mathbf{f}_2, \dots, \mathbf{f}_N\}$ be the temporally ordered set of $N$ flows, where each $\mathbf{f}_i \in \mathbb{R}^D$ is a $D$-dimensional feature vector. A temporal window $\mathbf{X}_t \in \mathbb{R}^{T \times D}$ is a sequence of $T$ consecutive flows. Each window $\mathbf{X}_t$ is assigned the ground-truth label of its terminal (most recent) flow, $y_t = y(\mathbf{f}_{(t+1) \cdot T - 1})$, formalizing a causal real-time monitoring setting where preceding historical traffic context informs the classification of the current network event.

A common weakness in NIDS evaluation is the use of overlapping windows. If a sliding window advances by a stride $S < T$, consecutive windows share $T-S$ flows. When the train/test boundary lies within these shared flows, test data leak into the training set.

To prevent this temporal leakage, we use non-overlapping windows with stride equal to the window length ($S = T$), and we set $T = 20$. Window length trades off two costs: longer windows increase the computation and memory needed by recurrent models, while shorter windows may miss multi-step attack behaviour.
We chose $T = 20$ empirically as a balance between these costs for the flow granularity of CSE-CIC-IDS-2018.
The $t$-th window is therefore

\begin{equation}
\mathbf{X}_t = \begin{bmatrix} \mathbf{f}_{t \cdot T} \\ \mathbf{f}_{t \cdot T + 1} \\ \vdots \\ \mathbf{f}_{(t+1) \cdot T - 1} \end{bmatrix}
\end{equation}

With $S = T = 20$, $\mathbf{X}_t \cap \mathbf{X}_{t'} = \emptyset$ for all $t \neq t'$, so no flow belongs to more than one window. Provided whole windows are assigned to a single split, no flow can appear in both the training and the test set.

\subsubsection{Spatial Edge Construction via $k$-NN}

A temporal window $\mathbf{X}_t$ records the chronological sequence of flows but does not capture latent behavioural correlations, such as flows originating from different source hosts that exhibit coordinated or statistically similar traffic patterns. Because the feature-extracted release of CSE-CIC-IDS-2018 used here provides no explicit network topology graphs or raw packet headers, we infer spatial relationships directly from the flow feature space using $k$-Nearest Neighbours ($k$-NN).

Crucially, graph generation is performed strictly on an intra-window basis: each individual network flow $\mathbf{f}_i \in \mathbb{R}^D$ is treated as a single node in the graph. Therefore, for every window of $T = 20$ consecutive flows, exactly one graph snapshot $\mathcal{G}_t = (\mathbf{X}_t, \mathbf{A}_t)$ is generated, consisting of $T = 20$ nodes and an intra-window adjacency matrix $\mathbf{A}_t \in \{0, 1\}^{20 \times 20}$. For each flow node $\mathbf{f}_i$ within window $t$, we compute the pairwise Euclidean distance to every other flow in the exact same window and connect $\mathbf{f}_i$ directed to its $k$ nearest neighbours in feature space. Edges are never constructed across different temporal windows; the graph topology is generated wholly within each 20-flow block, ensuring that no future flow information leaks into earlier snapshots and preserving strict temporal separation. The adjacency matrix $\mathbf{A}_t \in \{0, 1\}^{T \times T}$ (where $T = 20$) is formally defined as
\begin{equation}
A_{t, i, j} =
\begin{cases}
1, & \text{if } \mathbf{f}_j \in k\text{-NN}(\mathbf{f}_i) \\
0, & \text{otherwise}
\end{cases}
\end{equation}
We set $k = 5$. The resulting output of this construction is a sequence of independent graph snapshots $\mathcal{G} = \{\mathcal{G}_1, \mathcal{G}_2, \dots, \mathcal{G}_M\}$, where each individual snapshot $\mathcal{G}_t = (\mathbf{X}_t, \mathbf{A}_t)$ corresponds to exactly one 20-flow window and encapsulates both the ordered node feature representations and the inferred intra-window similarity topology. These snapshots serve as the input to ST-GAT.

% ------------------------------------------------------------
\subsection{Proposed ST-GAT Architecture}
% ------------------------------------------------------------

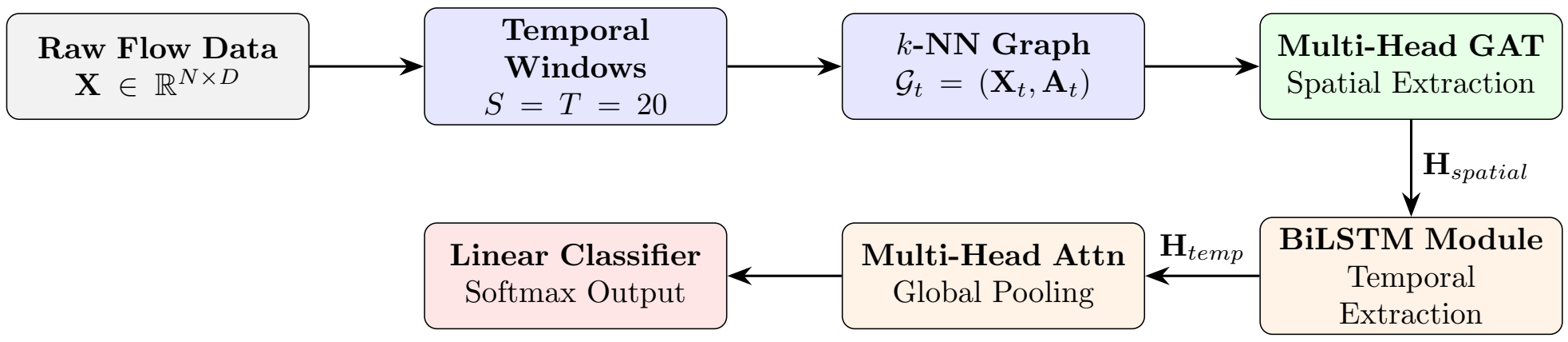
\begin{figure*}[htbp]
\centering
\resizebox{\linewidth}{!}{
\begin{tikzpicture}[
  node distance=1.1cm and 1.3cm,
  block/.style={rectangle, draw, fill=blue!10, text width=3.2cm, text centered, rounded corners, minimum height=1.2cm},
  arrow/.style={-{Stealth[scale=1.2]}, thick}
]

\node[block, fill=gray!10] (input) {\textbf{Raw Flow Data}\\$\mathbf{X} \in \mathbb{R}^{N \times D}$};
\node[block, right=of input] (window) {\textbf{Temporal Windows}\\$S = T = 20$};
\node[block, right=of window] (knn) {\textbf{$k$-NN Graph}\\$\mathcal{G}_t = (\mathbf{X}_t, \mathbf{A}_t)$};
\node[block, fill=green!10, right=of knn] (gat) {\textbf{Multi-Head GAT}\\Spatial Extraction};
\node[block, fill=orange!10, below=of gat] (bilstm) {\textbf{BiLSTM Module}\\Temporal Extraction};
\node[block, fill=orange!10, left=of bilstm] (attn) {\textbf{Multi-Head Attn}\\Global Pooling};
\node[block, fill=red!10, left=of attn] (output) {\textbf{Linear Classifier}\\Softmax Output};

\draw[arrow] (input) -- (window);
\draw[arrow] (window) -- (knn);
\draw[arrow] (knn) -- (gat);
\draw[arrow] (gat) -- (bilstm) node[midway, right] {$\mathbf{H}_{spatial}$};
\draw[arrow] (bilstm) -- (attn) node[midway, above] {$\mathbf{H}_{temp}$};
\draw[arrow] (attn) -- (output);

\end{tikzpicture}
}
\caption{Architecture of the proposed Spatio-Temporal Graph Attention Network (ST-GAT). Tabular flow data are partitioned into non-overlapping temporal windows ($T = 20$), where each flow forms a node and exactly one $k$-NN graph snapshot is constructed per window ($20$ nodes, $20 \times 20$ adjacency matrix). Multi-head GAT, a BiLSTM, and temporal attention produce a joint spatio-temporal representation for classification.}
\label{fig:architecture}
\end{figure*}

ST-GAT integrates graph-based spatial learning with recurrent temporal modelling in a single end-to-end model. As Fig.~\ref{fig:architecture} shows, it processes each graph snapshot in three stages: spatial feature extraction, temporal sequence aggregation, and classification.

\subsubsection{Spatial Feature Extraction (GAT Layer)}

The first module processes each graph snapshot $\mathcal{G}_t = (\mathbf{X}_t, \mathbf{A}_t)$ (comprising $T = 20$ flow nodes) independently with a multi-head Graph Attention Network (GAT), which aggregates information from each node's neighbours. A GCN weights neighbours by fixed, degree-based coefficients. GAT instead learns a separate coefficient $\alpha_{ij}$ for each edge from the features of both endpoints, so it can down-weight uninformative neighbours.

For node $\mathbf{f}_i$, the spatial representation concatenates the outputs of $K$ independent attention heads:
\begin{equation}
\mathbf{h}_{spatial, i} = \Big\|_{k=1}^{K} \sigma\left(\sum_{j \in \mathcal{N}(i)} \alpha_{ij}^k \mathbf{W}^k_{spatial} \mathbf{f}_j\right)
\end{equation}
where $\mathbf{W}^k_{spatial}$ is the weight matrix of head $k$, $\sigma$ is a non-linear activation (LeakyReLU), and $\alpha_{ij}^k$ is the normalised attention coefficient of head $k$ for the edge from flow $j$ to flow $i$. The rows of the resulting matrix $\mathbf{H}_{spatial} \in \mathbb{R}^{T \times D'}$ are spatially enriched flow embeddings, kept in temporal order.

\subsubsection{Temporal Sequence Modelling (BiLSTM Layer)}

To model how an attack evolves over time, the rows of $\mathbf{H}_{spatial}$ are fed in order to a BiLSTM. Unlike a unidirectional RNN, the BiLSTM reads the window both forwards and backwards in time.

At each time step $t \in \{1, \dots, T\}$, the BiLSTM outputs the concatenation of its forward and backward hidden states:
\begin{equation}
\mathbf{h}_{temp, t} = \overrightarrow{\text{LSTM}}(\mathbf{h}_{spatial, t}) \oplus \overleftarrow{\text{LSTM}}(\mathbf{h}_{spatial, t})
\end{equation}
where $\oplus$ denotes concatenation. The resulting sequence $\mathbf{H}_{temp} \in \mathbb{R}^{T \times D''}$ represents the window's spatio-temporal context.

\subsubsection{Global Temporal Pooling (Multi-Head Attention)}

Instead of taking the mean or the last hidden state of the BiLSTM, the model pools over time with multi-head attention. This lets the model learn which time steps (flows) in a window are most relevant to detecting an intrusion.

The sequence is projected into Query ($\mathbf{Q}$), Key ($\mathbf{K}$), and Value ($\mathbf{V}$) spaces, and a weighted context is computed as
\begin{equation}
\mathbf{C} = \text{softmax}\left(\frac{\mathbf{Q}\mathbf{K}^\top}{\sqrt{d_k}}\right)\mathbf{V}
\end{equation}
The context $\mathbf{C}$ is then flattened into a single vector that summarises the spatial and temporal structure of the 20-flow window.

\begin{algorithm}[htbp]
\caption{ST-GAT Forward Pass and Training Update}
\label{alg:stgat_training}
\begin{algorithmic}[1]
\renewcommand{\algorithmicrequire}{\textbf{Input:}}
\renewcommand{\algorithmicensure}{\textbf{Output:}}
\REQUIRE Raw network flows $\mathcal{F}$, sequence length $T=20$, $k$-NN neighbors $k=5$
\ENSURE Updated model weights $\mathbf{W}$
\STATE Remove undefined values and rare classes
\STATE Apply stratified train-test split
\STATE Fit \texttt{MinMaxScaler} on training set exclusively
\STATE Group flows into non-overlapping temporal windows $\mathbf{X}$ (stride $S = T$)
\FOR{\textbf{each} training batch \textbf{do}}
    \STATE Construct spatial graph $\mathcal{G} = (\mathbf{X}, \mathbf{A})$ using $k$-NN
    \STATE Compute spatial representations $\mathbf{H}_{spatial}$ via Multi-Head GAT
    \STATE Compute temporal features $\mathbf{H}_{temp}$ via BiLSTM
    \STATE Aggregate global context $\mathbf{C}$ via Temporal Attention
    \STATE Output class probabilities $\hat{y}$ via linear classifier
    \STATE Calculate Inverse-Weighted Cross Entropy Loss $\mathcal{L}_{WCE}$
    \STATE Backpropagate gradients and update $\mathbf{W}$ via Adam
\ENDFOR
\RETURN $\mathbf{W}$
\end{algorithmic}
\end{algorithm}

\subsubsection{Classification Head}

The pooled vector is passed to a fully connected layer followed by a softmax. The output is a probability distribution over the \texttt{Benign} class and the attack classes for the window.

\subsection{Handling Class Imbalance: Loss Reweighting vs. Resampling}\label{sec:imbalance}

As the EDA demonstrated, CSE-CIC-IDS-2018 exhibits extreme class imbalance, where benign background flows vastly outnumber intrusion attempts. Under standard unweighted cross-entropy, gradient descent is heavily dominated by the majority \texttt{Benign} class, leaving minority attacks under-penalized. While no algorithmic technique can artificially replace genuine attack diversity or fundamentally eliminate the statistical challenge of learning from scarce observations, detectors must adopt disciplined mechanisms to handle this skew without corrupting the feature space. Within ST-GAT, we compare three distinct strategies: cost-sensitive loss reweighting, batch-level balanced resampling, and prototype-based metric learning.

\subsubsection{Inverse Class-Frequency Weighting (Standard Baseline)}

The first strategy changes the objective function instead of the data distribution. The categorical cross-entropy (CE) loss is weighted by inverse class frequency, so errors on rare attack classes are penalised heavily and errors on common classes lightly. The weight of class $c$ is

\begin{equation}
w_c = \frac{N}{C \cdot N_c}
\end{equation}
where $N$ is the total number of temporal windows in the training set, $C$ is the number of classes, and $N_c$ is the number of windows of class $c$. The weighted loss for a prediction $\hat{y}$ with one-hot ground truth $y$ is
\begin{equation}
\mathcal{L}_{WCE} = - \sum_{c=1}^{C} w_c \cdot y_c \log(\hat{y}_c)
\end{equation}
With these weights, a misclassified \texttt{DoS attacks-Slowloris} window contributes to the loss roughly as much as several hundred misclassified \texttt{Benign} windows. The weight ratio $w_{\text{Slowloris}}/w_{\text{Benign}} = N_{\text{Benign}}/N_{\text{Slowloris}}$ is about 456, estimated from the stratified test supports in Table~\ref{tab:per_class_f1} (1,367/3).

\subsubsection{Dynamic Balanced Sampling (Oversampling)}

The second strategy changes the class distribution of the batches the network sees. Instead of drawing consecutive windows, we use PyTorch's \texttt{WeightedRandomSampler}, which samples window $i$ with probability $p_i$ proportional to the reciprocal of its class frequency.

The data loader therefore oversamples minority-class windows with replacement and heavily undersamples \texttt{Benign}. In expectation, each batch of $B=256$ windows contains roughly equal numbers from every class. Every attack type thus appears in every epoch, whereas without resampling the rarest classes would appear in very few batches.

\subsubsection{Few-Shot Metric Prototyping (MetricProto)}

The third strategy treats rare-attack detection as few-shot metric learning. ST-GAT embeddings are mapped into a metric space, and a prototype $\mathbf{p}_c \in \mathbb{R}^d$ is computed for each class $c$ as the mean embedding of that class's support samples.

Training minimises the Euclidean distance between a query window's embedding and its class prototype while maximising the distance to the other prototypes. Class probabilities are given by a softmax over negative squared distances:

\begin{equation}
P(\hat{y} = c \mid \mathbf{x}) = \frac{\exp(-||\text{ST-GAT}(\mathbf{x}) - \mathbf{p}_c||^2)}{\sum_{k=1}^{C} \exp(-||\text{ST-GAT}(\mathbf{x}) - \mathbf{p}_k||^2)}
\end{equation}
The intent is to form compact class clusters in the latent space that keep rare attacks apart from the dense \texttt{Benign} region seen in the PCA projection.

\section{Experimental Setup}\label{sec:setup}
% ============================================================

This section describes the computational environment, the training hyperparameters, and the evaluation metrics.

% ------------------------------------------------------------
\subsection{Environment and Hyperparameters}
% ------------------------------------------------------------

All models were implemented in PyTorch, and graph operations used the PyTorch Geometric (PyG) library. Training ran on an NVIDIA GPU.

ST-GAT and all deep baseline variants were trained with Adam at an initial learning rate of $0.001$, a batch size of $B = 256$, and at most $100$ epochs.

To limit overfitting, we used early stopping on validation Macro-F1. Training stopped when validation Macro-F1 had not improved for $20$ epochs (patience $= 20$), and the checkpoint with the best validation Macro-F1 was used for testing.

To estimate variability due to weight initialisation, every configuration was trained with three random seeds. Section~\ref{sec:results} reports the mean and standard deviation over these three runs.

% ------------------------------------------------------------
\subsection{Evaluation Metrics}
% ------------------------------------------------------------
Intrusion detection is a multi-class classification task. For each class $c$ we count true positives ($TP_c$), false positives ($FP_c$), and false negatives ($FN_c$) in a one-vs-rest fashion, and compute precision and recall:
\begin{equation}
\text{Precision}_c = \frac{TP_c}{TP_c + FP_c}, \quad \text{Recall}_c = \frac{TP_c}{TP_c + FN_c}
\end{equation}
The F1-score is the harmonic mean of precision and recall, so it penalises a model that trades one heavily for the other:
\begin{equation}
\text{F1}_c = 2 \cdot \frac{\text{Precision}_c \cdot \text{Recall}_c}{\text{Precision}_c + \text{Recall}_c}
\end{equation}

Accuracy (the fraction of correct predictions) is widely reported in the NIDS literature, but it is misleading on imbalanced data such as CSE-CIC-IDS-2018. On our test split, a classifier that always predicts \texttt{Benign} scores 73.7\% accuracy while detecting no attacks.

Our primary metric is therefore \textbf{Macro-F1}, the unweighted mean of per-class F1-scores over all $C$ classes:

\begin{equation}
\text{Macro-F1} = \frac{1}{C} \sum_{c=1}^{C} \text{F1}_c
\end{equation}

Because every class counts equally regardless of its support, Macro-F1 heavily penalises a model that fails on rare attacks such as \texttt{DoS attacks-Slowloris}. It is therefore a better measure of rare-attack detection than accuracy or weighted F1.

\section{Results and Discussion}\label{sec:results}
% ============================================================

This section compares ST-GAT with the baseline models. Unless stated otherwise, all figures are means and standard deviations over three independent random seeds.

% ------------------------------------------------------------
\subsection{Main Performance Analysis}
% ------------------------------------------------------------

We compared ST-GAT (trained with inverse class weights) with a Graph Convolutional Network variant (ST-GCN), with the same architecture trained using the other two imbalance strategies, and with per-flow MLP and Random Forest baselines. Table~\ref{tab:main_results} summarises the results.

\begin{figure}[htbp]
\centering
\includegraphics[width=\linewidth]{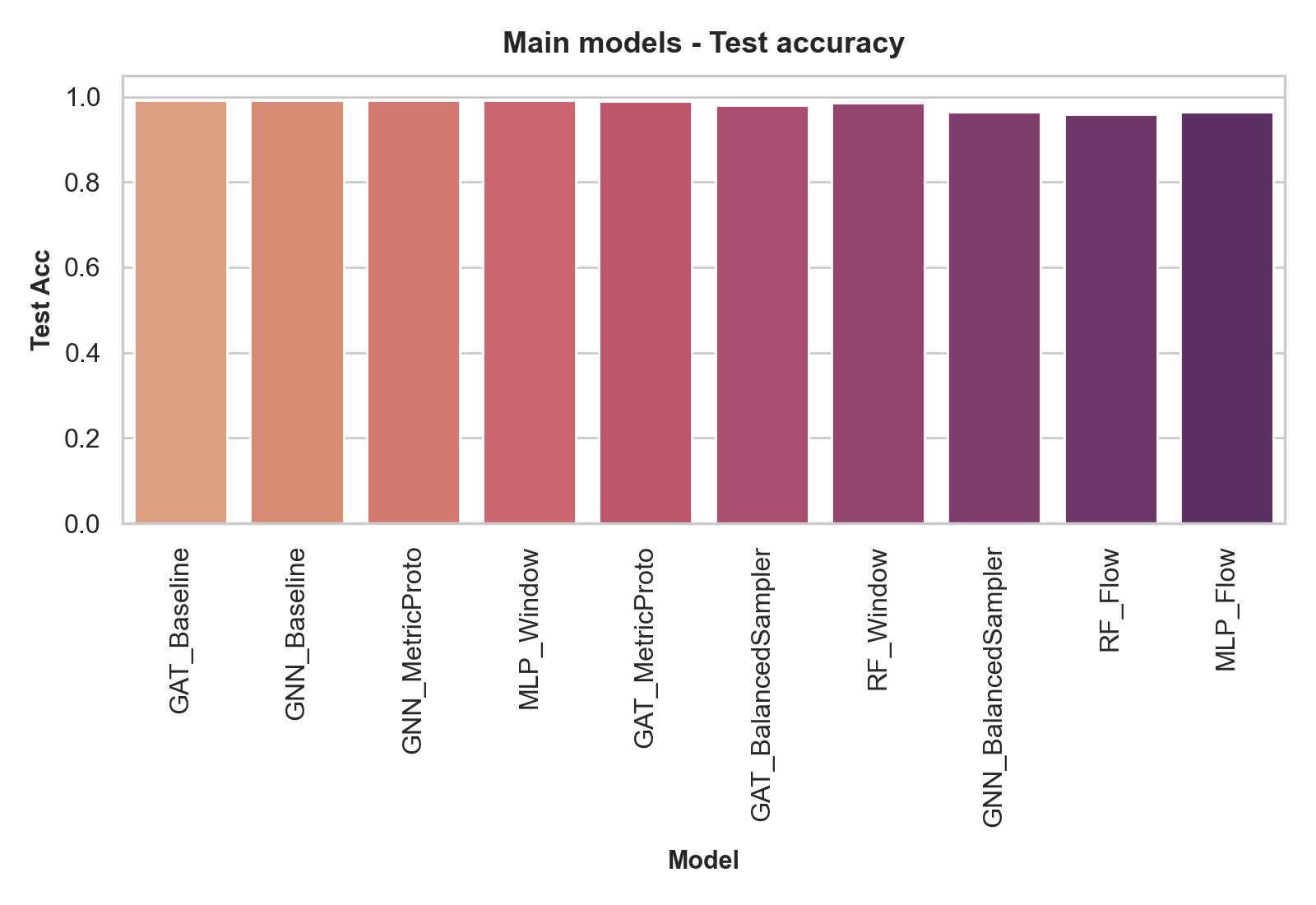}
\caption{Mean test accuracy of all evaluated architectures and sampling strategies.}
\label{fig:main_acc}
\end{figure}

\begin{figure}[htbp]
\centering
\includegraphics[width=\linewidth]{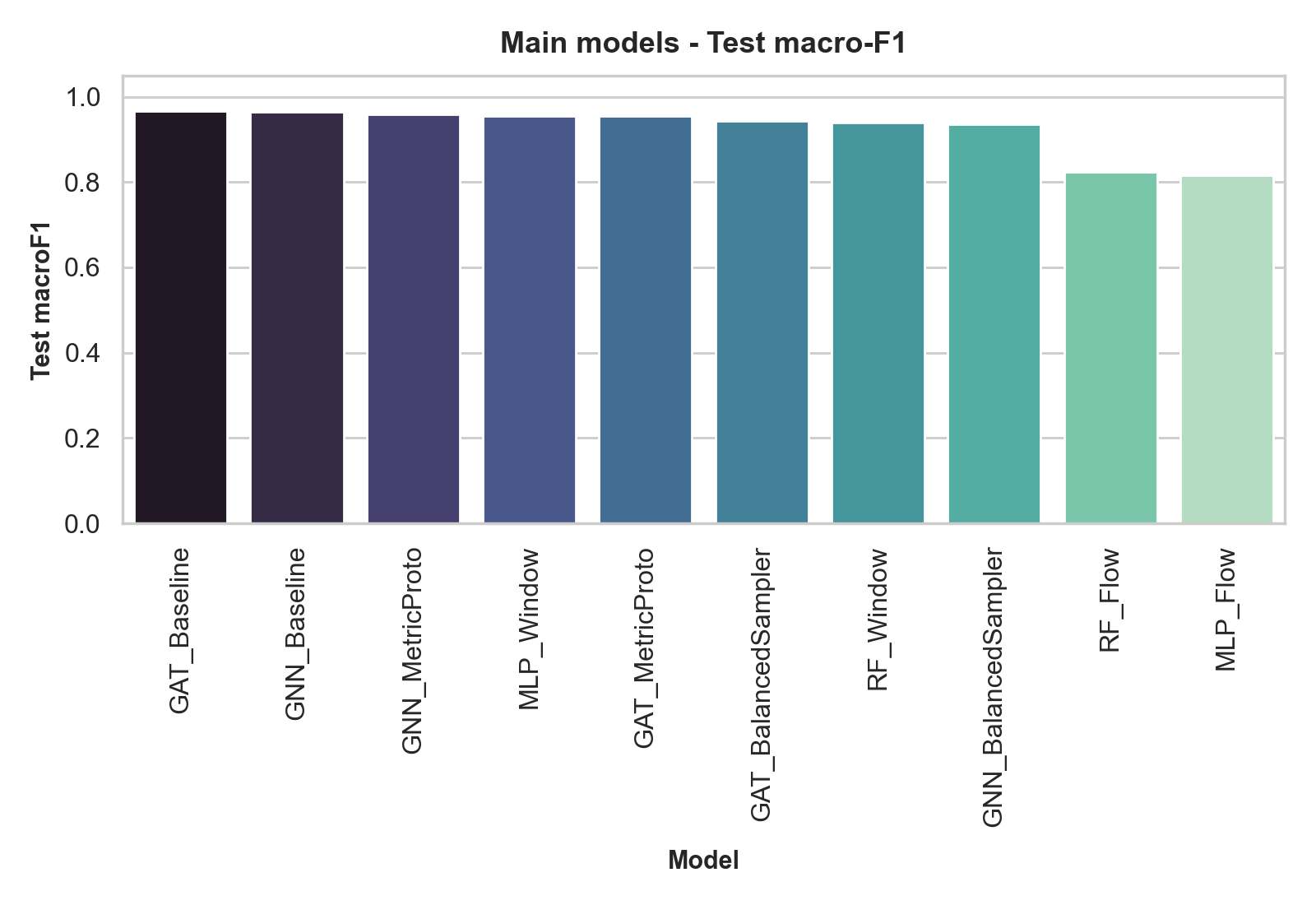}
\caption{Mean test Macro-F1 of all evaluated architectures. The window-level graph models exceed the per-flow classical baselines by about 14--15 points.}
\label{fig:main_f1}
\end{figure}

As Table~\ref{tab:main_results} and Figs.~\ref{fig:main_acc} and~\ref{fig:main_f1} show, ST-GAT achieved the highest mean Macro-F1 of all evaluated configurations, $\mathbf{0.9665 \pm 0.0087}$.

The per-flow baselines scored substantially lower on the minority attack classes. Evaluated on isolated flows under clean, uncorrupted feature distributions with cost-sensitive class weighting, Random Forest reached an accuracy of $0.9594 \pm 0.0003$ and a Macro-F1 of $0.8234 \pm 0.0054$, while the per-flow MLP reached an accuracy of $0.9646 \pm 0.0024$ and a Macro-F1 of $0.8167 \pm 0.0133$. When classical models were provided with window-level statistical aggregations across the 20-flow temporal window (mean and standard deviation pooling), performance improved to a Macro-F1 of $0.9387 \pm 0.0017$ (RF\_Window) and $0.9552 \pm 0.0013$ (MLP\_Window). However, the proposed ST-GAT model outperformed the per-flow Random Forest and MLP baselines by 14.3 and 15.0 percentage points in Macro-F1, respectively, and exceeded even the window-pooled MLP by 1.13 Macro-F1 points ($0.9665 \pm 0.0087$ vs.\ $0.9552 \pm 0.0013$). This demonstrates that explicit spatial graph attention over $k$-NN neighbourhoods combined with recurrent sequence modelling captures multi-step and coordinated attack behaviours significantly better than simple statistical feature pooling. The difference between ST-GAT and ST-GCN is only 0.25 points ($0.9665 \pm 0.0087$ vs.\ $0.9640 \pm 0.0057$), well within one standard deviation over three seeds.

\begin{table}[htbp]
\renewcommand{\arraystretch}{1.3}
\caption{Aggregate Performance of Main Models (mean $\pm$ std over 3 Seeds)}
\label{tab:main_results}
\centering
\begin{tabular}{l c c}
\toprule
\textbf{Model \& Strategy} & \textbf{Accuracy} & \textbf{Macro-F1} \\
\midrule
\textbf{ST-GAT (Proposed)} & $0.9914 \pm 0.0009$ & $\mathbf{0.9665 \pm 0.0087}$ \\
ST-GCN (Baseline) & $0.9905 \pm 0.0024$ & $0.9640 \pm 0.0057$ \\
\midrule
GAT\_MetricProto & $0.9896 \pm 0.0009$ & $0.9543 \pm 0.0027$ \\
GNN\_MetricProto & $\mathbf{0.9916 \pm 0.0013}$ & $0.9581 \pm 0.0035$ \\
\midrule
GAT\_BalancedSampler & $0.9799 \pm 0.0005$ & $0.9437 \pm 0.0145$ \\
GNN\_BalancedSampler & $0.9641 \pm 0.0165$ & $0.9345 \pm 0.0194$ \\
\midrule
RF\_Flow (Per-Flow) & $0.9594 \pm 0.0003$ & $0.8234 \pm 0.0054$ \\
MLP\_Flow (Per-Flow) & $0.9646 \pm 0.0024$ & $0.8167 \pm 0.0133$ \\
RF\_Window (Window-Pooled) & $0.9865 \pm 0.0004$ & $0.9387 \pm 0.0017$ \\
MLP\_Window (Window-Pooled) & $0.9907 \pm 0.0009$ & $0.9552 \pm 0.0013$ \\
\bottomrule
\end{tabular}
\end{table}

% ------------------------------------------------------------
\subsection{Analysis of Class Imbalance Strategies}\label{sec:imbalance_results}
% ------------------------------------------------------------

We compared the three imbalance strategies of Section~\ref{sec:imbalance} on both graph backbones. Table~\ref{tab:main_results} shows that, despite its simplicity, inverse class weighting outperformed both balanced sampling and few-shot prototyping.

\texttt{BalancedSampler} gave the lowest Macro-F1 on both backbones: $0.9437$ for GAT and $0.9345$ for the GCN variant, compared with $0.9665$ and $0.9640$ under inverse weighting. It also had the largest seed-to-seed variance. Repeatedly resampling the few available minority windows with replacement causes the model to overfit noisy minority artifacts, while undersampling substantially reduces the diversity of benign traffic the model encounters. \texttt{MetricProto} performed better ($0.9543$ with GAT and $0.9581$ with the GCN variant) but still below inverse weighting. These findings underscore that while algorithmic treatments do not eliminate the fundamental statistical challenge of scarce attack data, cost-sensitive loss reweighting provides an effective gradient stabilizer that manages class skew far more reliably than aggressive sample-space manipulation.

% ------------------------------------------------------------
\subsection{Architecture Ablation Study}\label{sec:ablation}
% ------------------------------------------------------------

\begin{figure}[htpb]
\centering
\includegraphics[width=0.85\linewidth]{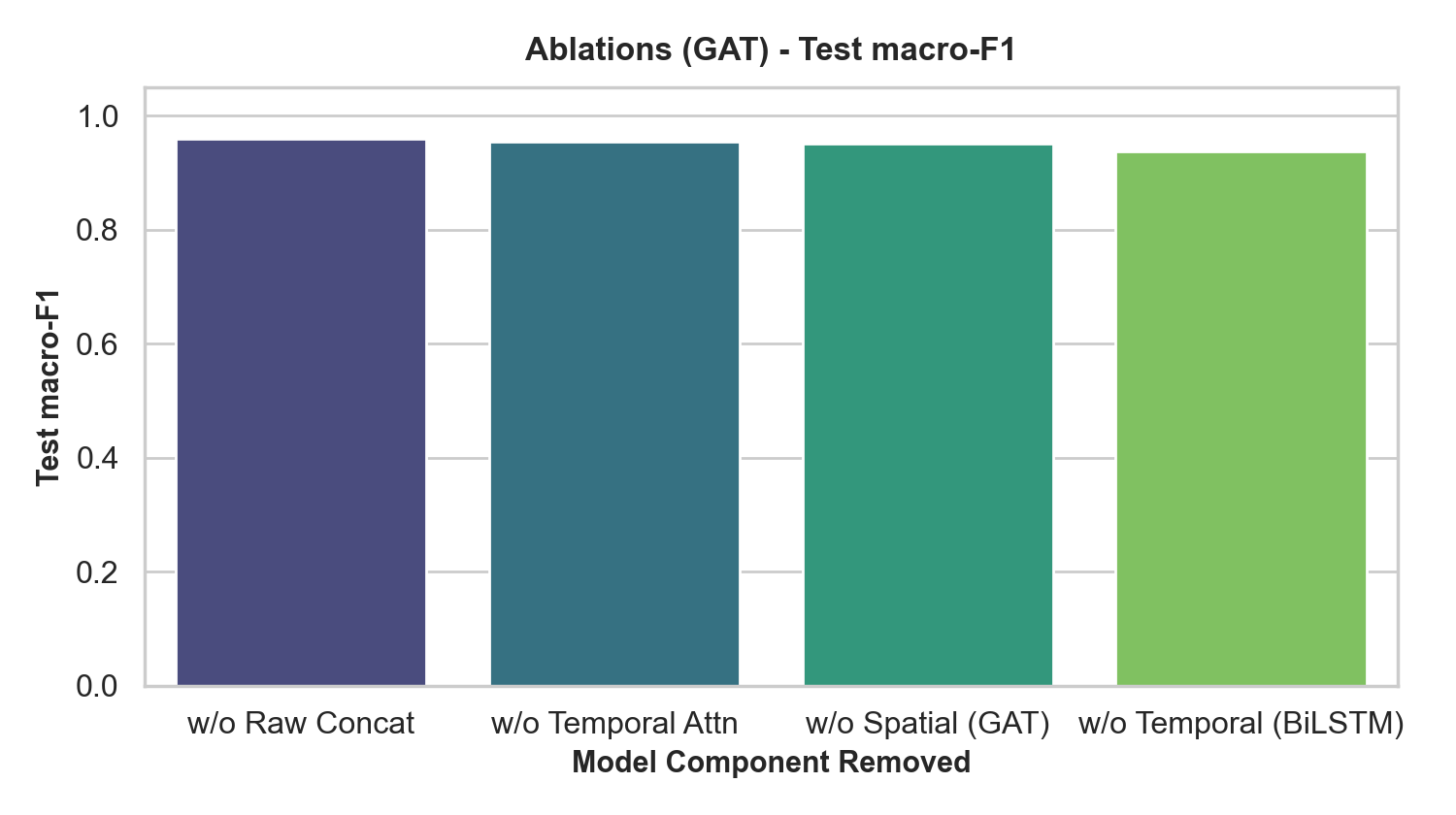}
\caption{Macro-F1 after removing individual architectural components. Removing the BiLSTM causes the largest drop.}
\label{fig:ablation_f1}
\end{figure}

To measure the contribution of each architectural component, we trained four ablated variants. Each removes one component: the raw-feature concatenation (\texttt{NoRawConcat}), the temporal attention pooling (\texttt{NoAttn}), the GAT spatial layer (\texttt{NoSpatial}), or the BiLSTM temporal layer (\texttt{NoLSTM}). Table~\ref{tab:ablation_results} and Fig.~\ref{fig:ablation_f1} report the results.

\begin{table}[htbp]
\renewcommand{\arraystretch}{1.3}
\caption{ST-GAT Architecture Ablation Study (mean $\pm$ std over 3 Seeds)}
\label{tab:ablation_results}
\centering
\begin{tabular}{l c c}
\toprule
\textbf{Ablated Architecture} & \textbf{Accuracy} & \textbf{Macro-F1} \\
\midrule
\textbf{ST-GAT (Full Model)} & $0.9914 \pm 0.0009$ & $\mathbf{0.9665 \pm 0.0087}$ \\
\midrule
w/o Raw Concat & $\mathbf{0.9921 \pm 0.0011}$ & $0.9591 \pm 0.0016$ \\
w/o Temporal Attn & $0.9892 \pm 0.0009$ & $0.9544 \pm 0.0017$ \\
w/o Spatial (GAT) & $0.9874 \pm 0.0026$ & $0.9503 \pm 0.0045$ \\
w/o Temporal (BiLSTM) & $0.9772 \pm 0.0065$ & $0.9365 \pm 0.0074$ \\
\bottomrule
\end{tabular}
\end{table}

Removing the BiLSTM (\texttt{NoLSTM}) caused the largest drop, lowering Macro-F1 from 0.9665 to $0.9365$ (3.0 points). This is consistent with the EDA observation that much of the feature variance lies in inter-arrival-time statistics, and it indicates that modelling flow order within a window is the most important component of ST-GAT on this dataset. Removing the GAT spatial layer (\texttt{NoSpatial}) lowered Macro-F1 to $0.9503$ (1.6 points), so the $k$-NN graph adds information beyond what the sequence model captures alone. Removing temporal attention or raw-feature concatenation cost 1.2 and 0.7 points, respectively. Removing raw-feature concatenation slightly increased mean accuracy (0.9921) while lowering Macro-F1. In other words, the raw features mainly help the rare classes.

% ------------------------------------------------------------
\subsection{Per-Class Analysis and Confusion Matrix}
% ------------------------------------------------------------

\begin{figure}[htbp]
\centering
\includegraphics[width=0.85\linewidth]{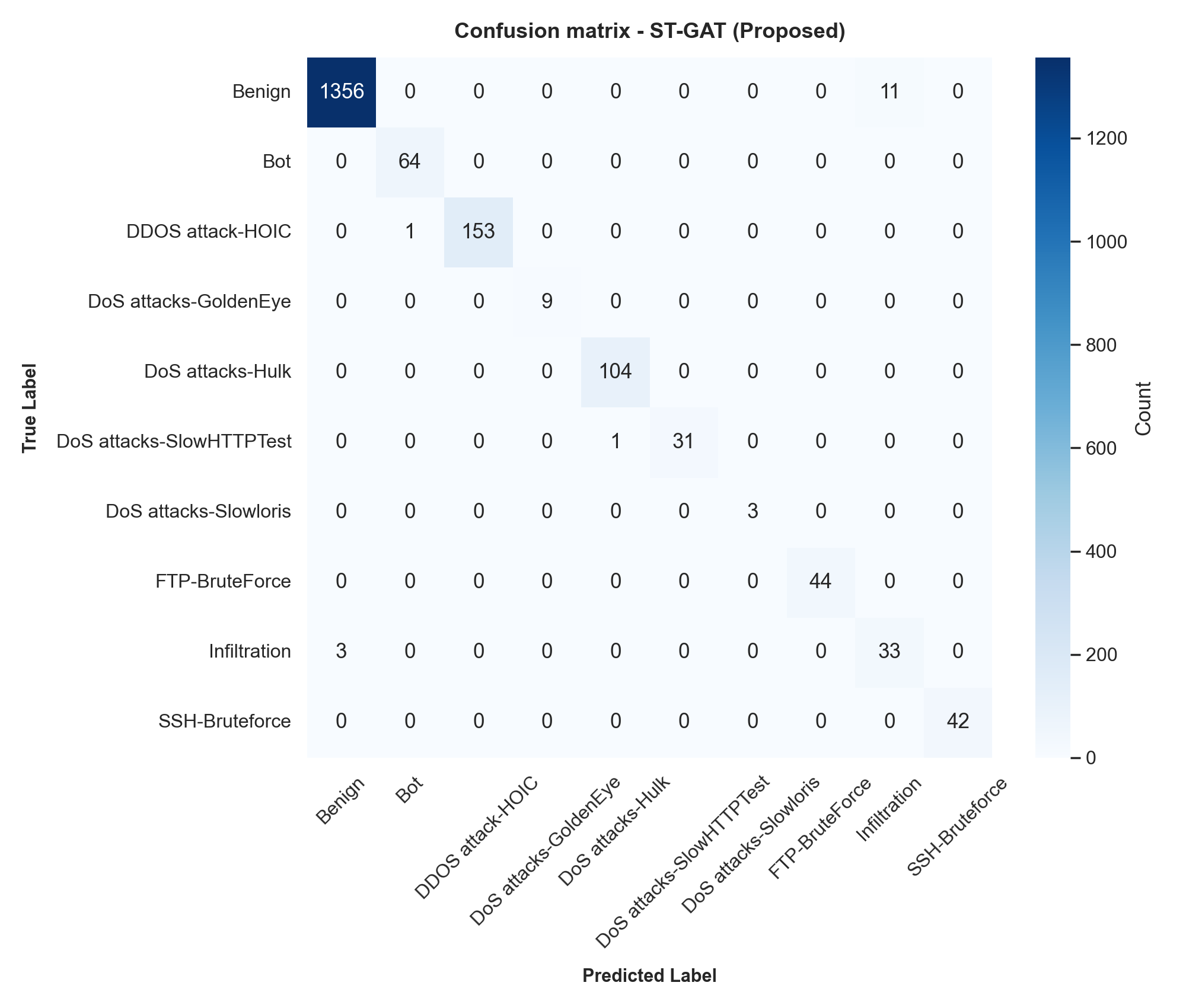}
\caption{Test-set confusion matrix of the best-performing ST-GAT run. Minority attacks are detected without collapsing into the Benign majority; the main error is Benign windows predicted as Infiltration.}
\label{fig:cm_gat}
\end{figure}

Macro-F1 summarises performance across classes, but it can hide weaknesses on individual rare classes. We therefore computed per-class precision, recall, and F1 from the test-set confusion matrix (Fig.~\ref{fig:cm_gat}) of the best-performing ST-GAT run. Table~\ref{tab:per_class_f1} lists the results.

\begin{table}[htbp]
\renewcommand{\arraystretch}{1.3}
\caption{Per-Class Performance of the Best-Performing ST-GAT Run (support = number of test windows)}
\label{tab:per_class_f1}
\centering
\begin{tabular}{l c c c c}
\toprule
\textbf{Class Label} & \textbf{Support} & \textbf{Precision} & \textbf{Recall} & \textbf{F1-Score} \\
\midrule
Benign & 1367 & 0.9978 & 0.9920 & 0.9949 \\
Bot & 64 & 0.9846 & 1.0000 & 0.9922 \\
DDOS attack-HOIC & 154 & 1.0000 & 0.9935 & 0.9967 \\
DoS attacks-GoldenEye & 9 & 1.0000 & 1.0000 & 1.0000 \\
DoS attacks-Hulk & 104 & 0.9905 & 1.0000 & 0.9952 \\
DoS attacks-SlowHTTPTest & 32 & 1.0000 & 0.9688 & 0.9841 \\
DoS attacks-Slowloris & 3 & 1.0000 & 1.0000 & 1.0000 \\
FTP-BruteForce & 44 & 1.0000 & 1.0000 & 1.0000 \\
SSH-Bruteforce & 42 & 1.0000 & 1.0000 & 1.0000 \\
Infiltration & 36 & 0.7500 & 0.9167 & 0.8250 \\
\bottomrule
\end{tabular}
\end{table}

Despite the heavy imbalance, ST-GAT with inverse class weighting achieved high precision and recall on most minority classes. It reached F1 $= 1.0000$ on \texttt{DoS attacks-GoldenEye}, \texttt{DoS attacks-Slowloris}, \texttt{FTP-BruteForce}, and \texttt{SSH-Bruteforce}. Importantly, these perfect scores rest on very small supports (3 and 9 test windows for Slowloris and GoldenEye), where even a single misclassification would substantially degrade performance. Therefore, while loss reweighting effectively maintains gradient sensitivity across rare attacks, these scores should be interpreted cautiously as reflecting robust sensitivity in this partition rather than proof that class scarcity has been fundamentally solved.

\texttt{Infiltration} was the only class with markedly lower performance (F1 $= 0.8250$). The confusion matrix (Fig.~\ref{fig:cm_gat}) shows that 11 \texttt{Benign} windows were classified as \texttt{Infiltration}, giving a precision of 0.75. Confusion between these two classes is common in NIDS evaluations. Infiltration attacks run from a compromised internal host and often generate traffic that is statistically similar to normal activity. Liu et al.~\cite{Liu2022} also report labelling problems affecting this class. Nevertheless, the model detected 33 of the 36 \texttt{Infiltration} windows (recall $= 0.9167$), so most infiltration activity would still raise an alert.

\section{Conclusion}\label{sec:conclusion}
% ============================================================

Reported NIDS performance is often inflated by methodological flaws, notably temporal leakage, pre-split contamination, and poor handling of minority classes. This paper presented ST-GAT, a spatio-temporal graph attention network designed and evaluated with these flaws in mind.

ST-GAT splits tabular flow data into disjoint time windows and infers spatial connections within each window with $k$-NN, so it models both the sequential and the relational structure of attacks. Among the evaluated strategies for handling class skew, inverse class-frequency weighting proved more reliable than balanced resampling and few-shot prototyping in sustaining Macro-F1, providing disciplined gradient stabilization without distorting empirical traffic distributions or overfitting scarce minority clusters.

On CSE-CIC-IDS-2018, ST-GAT achieved a mean test Macro-F1 of $\mathbf{0.9665}$ over three seeds, 14.3 points above a per-flow Random Forest baseline (0.8234). The ablation study identified the BiLSTM temporal component as the largest single contributor: removing it cost 3.0 Macro-F1 points.

Future work has three directions. First, the $k$-NN graph is a proxy for topology when endpoint identifiers are unavailable. We will test ST-GAT in settings where IP-level topology and routing structure are known, and compare feature-similarity graphs with host-topology graphs. Second, results on a single dataset cannot show that the learned patterns generalise. We will train on CSE-CIC-IDS-2018 and test on independent corpora such as UNSW-NB15 and ToN-IoT, which will show whether ST-GAT learns transferable intrusion patterns or artefacts of one testbed. Third, we will compare window lengths and labelling rules systematically, and evaluate on contiguous time-block splits to test robustness to distribution shift over time.

% ============================================================
% END OF MANUSCRIPT
% ============================================================

\backmatter

\section*{Declarations}
\begin{itemize}
\item \textbf{Funding:} Not applicable.
\item \textbf{Conflict of interest/Competing interests:} The authors declare that they have no competing interests.
\item \textbf{Availability of data and materials:} The dataset used is CSE-CIC-IDS-2018.
\item \textbf{Code availability:} Code will be made available upon reasonable request.
\item \textbf{Authors' contributions:} NP took the lead in conceptualizing the ST-GAT model, conducting the primary experiments, and writing the manuscript. SaP, SI, SiP, and SG contributed equally to data analysis, pipeline development, and reviewing the manuscript.
\end{itemize}

\bibliography{references}

\end{document}